\documentclass[12pt]{article}
\usepackage{epsfig,graphicx}

\title{Analysis of the QCD spectrum and  chiral symmetry breaking   with   varying  quark masses}
\author{ Yu.A.Simonov\\
State Research
Center\\Institute of Theoretical and Experimental Physics, \\
Moscow, 117218 Russia}

\newcommand{\beq}{\begin{eqnarray}}
 \newcommand{\eeq}{\end{eqnarray}}
\newcommand{\be}{\begin{equation}}
 \newcommand{\ee}{\end{equation}}

 \def\la{\mathrel{\mathpalette\fun <}}

\def\fun#1#2{\lower3.6pt\vbox{\baselineskip0pt\lineskip.9pt
\ialign{$\mathsurround=0pt#1\hfil ##\hfil$\crcr#2\crcr\sim\crcr}}}

\newcommand{{\SD}}{\rm SD}

\newcommand{\vex}{\mbox{\boldmath${\rm x}$}}

\newcommand{\vep}{\mbox{\boldmath${\rm p}$}}

\newcommand{\ves}{\mbox{\boldmath${\rm s}$}}

\newcommand{\lan}{\langle}
\newcommand{\ran}{\rangle}

\begin{document}
\maketitle
\begin{abstract}

The meson spectrum of QCD is studied in the framework of nonperturbative QCD
as a function of varying
 quark masses $m_q$. It is shown, that  the total spectrum consists of two branches:
1) the standard one, which may be called the flux-tube spectrum, depending
approximately  linearly on $m_q$ and 2) the chiral symmetry breaking (CSB)
spectrum for pseudoscalar flavor nonsinglet  (PS) mesons, with mass dependence
$\sqrt{m_q}$. The formalism for PS mesons is derived from  the QCD Lagrangian
with $m_q$ corrections, and a unified form of the PS propagator was derived. It
is shown, that the  CSB branch of PS mesons joins to the flux-tube branch at
around $m_q=200$ MeV. All these results are in close correspondence with recent
numerical data on large lattices.

\end{abstract}

\section{Introduction}
The connection   between the standard QCD picture of confinement, which may be called the flux-tube picture,and the CSB properties, which can be
shortly called the chiral physics, is not clearly understood by
physical community, especially on the level of model building.

The chiral Lagrangians have been introduced  before  the QCD era  to implement
CSB
  in \cite{1,2,3} and the GMOR relation \cite{4} contains an important
connection between pion characteristics $m_\pi,f_\pi$ and purely quark
characteristics $m_q$ and $\lan \bar q q\ran$. From the derivation it is not
clear, how this connection will transform with growing $m_q$, and in general at
what $m_q$ chiral mesons cease to be chiral and become standard flux-tube
mesons, which do not contain chiral effects,  since at large $m_q$ CSB is not
spontateous any more.On the side of chiral Lagrangians an effective technic of
Chiral Perturbation Theory (CPTh)\cite{5,e} was developed, which allows to
calculate all corrections in terms of additional terms, depending on
$m_\pi,f_\pi$. But one of the questions is that correction terms containing
$m_q$ can occur already in derivation of GMOR, which was actually known only in
the chiral limit. This will be demonstrated below explicitly.

To derive GMOR and explicit expressions for $m_\pi, f_\pi$ etc. we are using
the technic of derivation of chiral degrees of freedom in QCD, developed before
in \cite{a,b,c,d}. There it is shown explicitly how pionic variables appear in
QCD and how they are connected
 to the quark Green's functions. As a result all chiral dynamics can be expressed in terms of nonchiral (standard flux-tube)
quark-aniquark PS Green's function $G^{(0)}(\vep)$ and the so-called vertex (or residual) mass $M(0)$. Both these
quantities are calculated via string tension $\sigma$ and hence one can express chiral dynamics from the  first principles.

Another important point is the incorporation of the chiral branch,  i.e.
$\pi,K,\eta$ and the corresponding radial excited states into the general
scheme of   mesons, or in other words, to which extent  radial excitations of
$\pi, K$, e.g. $\pi (2S), K(2S)$ etc. are chiral objects.

Finally, studying meson spectra with varying $m_q$ one can check the
relativistic properties  of the Hamiltonian   $vs$ explicit numerical data and
discover interesting new dependencies. Indeed, as will be shown  below, the
spectrum of vector  mesons depends approximately  linearly on $m_q$ with good
accuracy both in analytic and lattice  calculations,  while for PS mesons
linear dependence for $m_q>250$ MeV (flux-tube part of spectrum) goes over into
the chiral $\sqrt{m_q}$ dependence for lower $m_q$,

To do all analytic calculations we are using the so-called Relativistic String
Hamiltonian (RSH) \cite{6}, which contains only
 first-principle parameters (no constituent masses or subtraction  constants) and was used successively for all hadrons: mesons
\cite{7}, baryons \cite{8}, glueballs \cite{9}, hybrids \cite{10}.

Good agreement in all cases with experiment and lattice data allows to exploit
RSH below for a precision test of flux-tube
 and chiral dynamics in QCD with varying $m_q$.

The plan of the  paper is as follows: in section 2 we shortly remind the
derivation of GMOR and expressions for $m_\pi, f_\pi$ in the context of QCD,
and find additional terms in $m_q$, which are important for growing $m_q$. We
also  compare  these results with lattice data.

In section 3  we calculate the flux-tube spectra of vector and PS mesons for
the whole set of $m_q$ values and compare those in detail with lattice data. We
find here linear dependencies of flux-tube spectra.

In section 4  we show how the  chiral spectrum is incorporated in the total spectrum  and what are chiral corrections to excited pions
and kaons. Section 5 contains summary and conclusions.

\section{Quark mass dependence of chiral dynamics}

In the standard approach one considers chiral Lagrangian and chiral
perturbation theory as a selfcontained approach and therefore all  quark mass
dependence (QMD) is usually derived  from chiral perturbation theory.
 In what follows we shall derive chiral  dynamics from QCD in a more direct
 way, as it was done in \cite{a,b,c,d}, and in addition we shall calculate
 correction terms and then define QMD for all basic quantities of chiral
 dynamics: $m_\pi, f_\pi $ and $\lan \bar \psi
 \psi\ran$.

 The starting point is the effective quark-meson Lagrangian, derived in {\cite{b},
 where both quark and chiral meson degrees of freedom (d.o.f) are kept,

 $$
 L_{QML} =\int d^4x\int d^4 y \{ ~^f\psi^+_{a\alpha} (x)[ (i\hat
   \partial + i m_f)_{\alpha\beta} \delta^{(4)}(x-y) \delta_{fg}
   +$$
   \be
   i M \hat U_{\alpha\beta}^{(fg)}(x,y)]~^g\psi_{a\beta}(y)- 2n_f[J(x,y)]^{-1}M^2(x,y) \}.
\label{1}\ee
 Here $\hat U^{(fg)}_{\alpha\beta}$ contains Nambu-Goldstone (NG) fields
 $\phi_a, \hat  U =\exp (i\gamma^5 \phi_a t^a),$ while $M(x,y)$ is the
 auxiliary scalar field. $J(x,y)$ is expressed as an integral of the confining
 field correlator $D(x)$, its exact form will not be used below.

After integrating quark d.o.f. one obtains the Effective Chiral Lagrangian
$L_{ECL}$, containing  effective scalar mass  variable  $M$ and $NG$ field
$\phi_a$, with the partition function \be Z=\int D MD\phi_a \exp
L_{ECL}\label{2}\ee where  $L_{ECL}$ is \be L_{ECL} =-2n_f\int d^4x \int d^4y
(J(x,y))^{-1}M^2(x,y) +N_c tr\log [(i\hat
\partial +im_f)\hat 1+ iM\hat U].\label{3} \ee

In (\ref{1}) and (\ref{3}) $M(x,y)$ enters  in its nonlocal form, the
nonlocality being of the order of the  vacuum correlation length
$\lambda\approx 0.1$ fm. In what follows only the local limit of $M(x,y) \to
M(x)$ will be used. In Fig1. one can see, that $M(x)$ can be associated with
the part of the flux tube, from the quark position $x$ to  the center of flux
tube; the same is true for antiquark.

\begin{figure}\begin{center}
\includegraphics[width= 5cm,height=6cm,keepaspectratio=true]{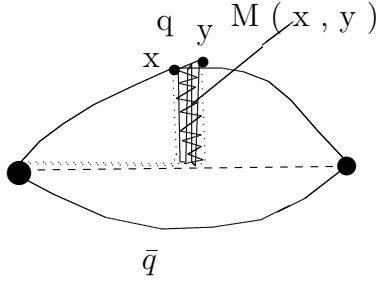}
\caption{The flux-tube operator $M(x,y)$ in the Wilson loop} \vspace{1cm}
\end{center}

\end{figure}

Expanding the last term in (\ref{3}) in powers of $\phi_a$, one obtains the
quadratic vertex in action \be W^{(2)}(\phi) =\frac{N_c}{2} \int \phi_a (k)
\phi_a(-k) \bar N(k) \frac{d^{(4)}k}{(2\pi)^4}\label{4}\ee where  \be \bar
N(k)= \frac12 tr\{(\Lambda M)_0+ \int d^{(4)}ze^{ikz}\Lambda(0,z) M(0) \bar
\Lambda(z,0) M(0)\}.\label{5}\ee

The graphical representation of $\bar N(k)$ is given in Fig.2.

\begin{figure}
\includegraphics[width= 9cm,height=10cm,keepaspectratio=true]{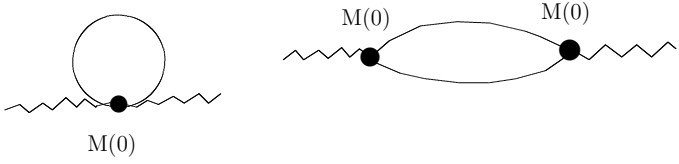}
\caption{Quadratic terms in the pionic action (5)} \vspace{1cm}

\end{figure}

 Here $\phi_a(k)$
is \be \phi_a t_a = \frac{\varphi_a \lambda_a}{f_\pi} =
\frac{\sqrt{2}}{f_\pi}\left(
\begin{array}{lll} \frac{\eta}{\sqrt{6}}+
\frac{\pi^o}{\sqrt{2}}, &\pi^+,& K^+\\
\pi^-, &\frac{\eta}{\sqrt{6}}-\frac{\pi^o}{\sqrt{2}},& K^o\\K^-,& \bar K^o,&
-\frac{2\eta}{\sqrt{6}}\\\end{array}\right)\label{6}\ee and $\Lambda, \bar
\Lambda$ are defined as

\be \Lambda= (\hat \partial + m + M)^{-1},~~ \bar \Lambda = (\hat
\partial - m-M)^{-1}.\label{7}\ee
\be \bar N(k) = \frac12 tr (\Lambda M \bar \Lambda (\hat\partial - m)) =
\frac12 [tr (\Lambda M)_0 + G^{(MM)} (k)]\label{8}\ee and the last term in
(\ref{8}) is \be G^{(MM)} (k) = tr (\Lambda M \bar \Lambda M)_k= G^{(MM)} (k) -
G^{(MM)} (0) + G^{(MM)}(0).\label{9}\ee Here and in (\ref{8}) the subscript
($o,k)$ indicates the momentum argument. For $N(0)$ one obtains from (\ref{8})
\be \bar N(0) = \frac12 (tr (\Lambda m) + tr (\Lambda m \bar \Lambda
m)).\label{10}\ee

Note the last term on the r.h.s. of (\ref{10}), which was not calculated before
in \cite{a,b,c,d}. Here $m$ is the diagonal quark matrix $ m= {\rm diag} (m_u,
m_d, m_s)$, and we shall confine ourselves to the $SU(2)$ case, replacing $m$
by $\frac{m_u+m_d}{2}$.

Now $tr \Lambda$ can be expressed via the standard chiral condensate, defined
in the Minkowskian space-time \be \lan \bar \psi \psi\ran_M = i \lan \psi
\psi^+\ran_E =-N_c tr \Lambda=\lan\bar u u + \bar dd\ran_M.\label{11}\ee To
trace down the QMD of $tr \Lambda$ it is convenient to rewrite it as \be tr
\Lambda = \lan tr (\gamma_5 \Lambda(x,y) \gamma_5 (M(0) +m) \Lambda (y,x) \ran
d^4y = - (M (0) +m ) G^{(0)} (k=0).\label{12}\ee

In (\ref{12}) and earlier in (\ref{8}) we have defined the Green's functions
$G^{(0)}$ and $G^{(MM)}$, which differ due to presence in the latter of the
operator $M$ at the initial and final points $x$ and $y$. In   a  general
position $M(x)$ contains confinement interaction of the given quark with
antiquark, i.e. it  exemplifies the string between $q$ and $\bar q$, see Fig.
1. However at the initial or the end point of the meson propagator it can be
expressed in terms of vacuum correlation length $\lambda$ \cite{15*},
$\lambda\approx 0.15$ fm, namely in the Appendix 3 of \cite{c} $M (0)$ was
estimated as \be M (0) = \frac{2}{\sqrt{\pi}} \sigma \lambda (1 +O(\sigma
\lambda^2)).\label{13}\ee
 As shown in \cite{b,c,d} $G^{(0)}$ and $G^{(MM)}$ can be represented as   spectral
 sums over meson states\be G^{(0)}(k) = -\sum _n \frac{c^2_n}{k^2+m^2_n}, ~~
 G^{(MM)}  (k) =- M^2(0) \sum \frac{c^2_n}{k^2+m^2_n},\label{14}\ee
 where $c_n= \sqrt{\frac{m_n}{2}}\psi_n(0)$, and $\psi_n(\vex)$ is the $n$-th
 state meson wave function. As a result one obtains for $\lan \bar
 \psi\psi\ran$,
 \be |\lan \bar \psi \psi \ran | = N_c  {(M (0) + m)} \sum^\infty_{n=0}
 \frac{|\psi_n (0)|^2}{ m_n}.\label{15}\ee
 For $\bar N(0)$ from (\ref{10}), (\ref{12}) one observes a cancellation of
 $O(m^2)$ correction terms,
 \be \bar N(0) = - \frac{mM(0)}{2} G^{(0)} (0) = \frac{mM(0)}{2}
 \sum^\infty_{n=0} \frac{|\psi_n(0)|^2}{2m_n}=\frac{mM(0)}{M(0)+m}
 \frac{|\lan \bar \psi\psi\ran|}{2N_c}\label{16}\ee

On the other hand, from the definition of $N(k)$ in (\ref{4}), one has a
correspondence

\be \bar N (k) = (m^2_\pi + k^2) \frac{f^2_\pi}{2N_c} + O(k^4)\label{17}\ee and
as a consequence the (modified)  GMOR relation \be m^2_\pi f_\pi^2 =
\frac{mM(0)}{M (0) +m } |\lan \bar \psi \psi\ran|,\label{18}\ee which goes over
into the standard GMOR formula in the chiral limit $m\to 0$ To find $f^2_\pi$
separately one  can use (\ref{8}), (\ref{9}) to write \be G^{(MM)} (k) -
G^{(MM)} (0) = \frac{k^2 f^2_\pi}{2N_c} + O(k^4)\label{19}\ee and finally \be
f^2_\pi= N_c M^2 (0) \sum_{n=0}^\infty
\frac{|\psi_n(0)|^2}{m^3_n}.\label{20}\ee

 Eqs. (\ref{15}),(\ref{20}) are exploited in the Appendix to obtain numerical
 values of $\lan \bar \psi \psi \ran$ and $f_\pi$ from the  calculated
 spectrum.

 Eqs. (\ref{15}),(\ref{18}),(\ref{20})
contain all information about QMD in the corresponding  quantities: to make it
explicit  one needs to tell about QMD  of  $m_n$ and  $\psi_n(0)$. First of all
one should mention, as in \cite{c}, that all sums over $n$ in (\ref{15}),
(\ref{16}), (\ref{20}) are formally divergent, since spectral representation
fails at times and distances less than $\lambda$, and one should use the
cut-off factor in integration over small time region, which effectively
produces the cut-off factors $e^{-m_n\lambda}$ in summation over $n$ in
(\ref{15}) and (\ref{20}).

This fact, however, does not influence  qualitatively the QMD of the
corresponding sums over $n$, which is mostly   contained in the first terms of
these sums, and appears there via $m_n$ and $\psi_n(0$, \be |\psi_n (0)|^2 =
\frac{\omega_n}{4\pi} \left(\sigma + \frac43 \alpha_s
\lan\frac{1}{r^2}\ran\right)\label{21}\ee where
$\omega_n=\lan\sqrt{p^2+m^2}\ran_n$.

As will be shown in the next section by analytic calculations and  comparison
to lattice data, the QMD of $m_n, \omega_n$ is rather mild, i.e. $m_n$ grows by
 $<20\%$ when $m_Q$ is changing from zero to 200 MeV, and in any case this
 dependence can be computed explicitly. Neglecting it for a moment, one expects
 with this accuracy:

 a)  A linear growth of $|\bar \psi \psi|$ with $m_q$,  i.e. for $m_q\approx M
 (0)\approx
0.15$ GeV the chiral condensate is twice as large as    compared to zero $m_q$
value.

b) Approximate independence of $f_\pi$ on $m_q$ for $m_q<0.2$ GeV in accordance
with Eq. (\ref{20}).

c) Approximate scaling  law $m_{PS}\sim \sqrt{m_q}$ for $m_q\la 0.2$ GeV.

More explicitly, $m_n$ and $\psi_n(0)$  depend on $m^2_q$, one can expect the following expansions at small $m_q$

\be m^2_{PS} =X m_q \left (1+O(m^2_q)\right),\label{22}\ee

\be f_{PS}(m_q) = f_{PS} (0) + O(m^2_q), \label{23}\ee

\be \lan \bar\psi\psi\ran_{m_q} = \lan\bar\psi\psi\ran_0 \left(
\frac{m_q+M(0)}{M(0)}\right) (1+O(m^2_q)).\label{24} \ee

This should be compared with  CPTh  results \cite{e}

\be m^2_{PS}= X m_q +Y m_q ln m_q +Z m^2_q\label{25}\ee

\be f_{PS} = f_0 (1+am^2_{PS})+ O(m^4_{PS})\label{26}\ee

Note, that CPTh  produces  additional terms with   respect  to our expansions
(\ref{22})-(\ref{24}), which originate from higher order terms in expansion of
$L_{ECL}$ in powers of $\phi^2_a$, not  accounted in our consideration above.

We can now compare our expansions (\ref{22}-\ref{24}) with lattice data,
obtained on large lattices  $ 16^332$ and $24^348$ \cite{f}. The detailed
comparison in \cite{f} of CPTh expansions (\ref{25}), (\ref{26}) with data has
shown a good agreement, which supports also our results (\ref{22}), (\ref{23}),
where no serious deviations from (\ref{25}), (\ref{26}) on the basis of proper
quark contributions is suggested.

The situation with the chiral condensate in quite different. Indeed, Eqs.
(\ref{18}) and (\ref{24}) show, that $\lan \bar \psi \psi\ran_0$, entering in
GMOR relation and $\lan \bar \psi \psi\ran$ at arbitrary value of  $m_q$,
differ by a linear factor  $K\equiv \frac{m_q +M(0)}{M(0)},$ which  can be
large, since $M(0) \approx \sigma \lambda \approx O(0.1$ GeV) (see Appendix 1
of \cite{c}). Additional corrections in $m_q$, entering from the sum in
(\ref{15}) are small for small $m_q< 0.2$ GeV, and the factor $K$ has a clear
significance in comparison to lattice data and experiment.

In CPTh the chiral condensate is difficult to treat because of divergence at
small $m_\pi$. On lattice the condensate $\lan \bar q q\ran$ was studied in
numerous papers, see  e.g. \cite{g}, \cite{h} and \cite{i} for a review and
references. Both in {\cite{g} and \cite{h} the    linear growth of $\lan \bar
\psi \psi\ran$ with $m_q$ was  also observed. As one can see from (\ref{15}),
(\ref{20}), $f_\pi\sim M(0) \sim \sigma \lambda$, while $\lan\bar q  q\ran \sim
\sigma M(0)\sim \sigma^2 \lambda$, where we have taken into account, that
$\psi^2_n (0) \sim \omega \sigma  \approx \sigma ^{3/2}, m_n \sim
\sqrt{\sigma}$. Inserting these estimates into the GMOR relation (\ref{18}),
one obtains

\be m^2_\pi \sim  m_q /\lambda, ~~ m_\pi \approx \sqrt{m_q}
\sqrt{M_G},\label{27a}\ee where  $M_G\approx 1/\lambda$ is the gluelump mass,
obtained in \cite{15*}, $M_G \approx 6\sqrt{\sigma} \approx 2$ GeV.

\section{The flux-tube   spectrum:  $ m_{ps} $ and $ m_v$}

In the previous section the use was made of the spectrum of states $\{m_n\}$
and  $\{ \psi_n (0)\}$, which correspond to a completely different regime,
described by the so-called Relativistic String Hamiltonian, which was derived
from the path-integral representation of the Green's function of quark and
antiquark at the ends of the QCD string. Therefore it  exemplifies both string
asymptotics at large $L$ and relativistic potential dynamics at small $L$. For
$L=0$ the RSH for the $q\bar q$ system has the form \be H=\sum^2_{i=1}
\sqrt{p^2+m_i} + V_\sigma (r) + V_g (r) + \Delta_{SE} + H_{ss},\label{35}\ee
where $V_\sigma (r) =\sigma r, ~~ V_g (r) =-\frac{4\alpha_B (r)}{3 r}$ , and
$\Delta_{SE}, H_{ss}$ are self-energy and hyperfine contributions to be defined
later.

 We shall exploit the so-called einbein version of RSH, which yields results
 numerically close to (\ref{35}),  but is easier to treat. In this case
  one introduces the auxiliary  variable $\omega_i$ and the total mass of the
  radial excited state $n, n=0,1,2,... $ can be written   for equal quark
  masses $m_1 =m_2\equiv m_q, ~\omega_1=\omega_2 =\omega,$ as

  \be M_n (\omega) = \frac{m^2_q}{\omega} + \omega +\varepsilon_n (\omega)
  +\Delta_{SE} (\omega) + M_{ss} (\omega)\equiv M^{(0)}_n (\omega) + \Delta_{SE} (\omega) + M_{ss}(\omega)\label{36}\ee
  with

$$\varepsilon_n (\omega) = \omega^{-1/3} \sigma^{2/3} a_n(\tau )~~ a_0{(0)}=2.338;~~\tau = \frac{4\alpha_s}{3}\left(\frac{\omega^2}{\sigma}
\right)^{1/3}$$

\be \Delta_{SE}=-\frac{3\sigma}{\pi\omega} \eta (m_q \lambda),~~M_{ss} (\omega)
=\frac{8 \alpha_{hf}}{9\omega^2} R^2_n (0) (\ves_1\ves_2), \label{37}\ee where
$\eta$ is a calculable  function, given  in \cite{i}
 note, that $\eta(0) =1$.

In the einbein method the equilibrium point of $M^{(0)}_n(\omega)$ is defined
by the equation

\be \frac{dM^{(0)}_n(\omega)}{d\omega}|_{\omega=\omega_0}=0,\label{38}\ee and
$\omega_0$ acquires the physical meaning of an average quark energy,
$\omega_0^{(n)}= \lan \sqrt{m^2_q+\vep^2}\ran_n$. Note, that all correction
terms, $\Delta_{SE} $ and $M_{ss}$, also depend on $\omega$, and we take them
in the first approximation at the point $\omega=\omega_0$.

In this way one obtains an equation for $\omega_0$,

\be \omega^2_0 = m^2_q +\frac{(\sigma\omega_0)^{2/3}}{3} a_n
(\tau)\label{39}\ee with solutions  which can be written in two forms, the
first appropriate for large $m_q$,

\be \omega^2_0 (n) =\sigma\left[ \left(
\frac{q}{2}+\sqrt{-\frac{p^3}{27}+\frac{q^2}{4}}\right)^{1/3}+
\left(\frac{q}{2}-
\sqrt{-\frac{p^3}{27}+\frac{q^2}{4}}\right)^{1/3}\right],\label{40}\ee while
the second is valid for small $m_q$, when $\frac{q^2}{4}< \frac{p^3}{27}$,

\be \omega^2_0 (n) = \sigma \left[ 2 \sqrt{\frac{p}{3}}\cos \left(\frac13
\arctan \left(\frac{2\sqrt{\frac{p^3}{27}-\frac{q^2}{4}}}{q}\right)
\right)\right]\label{41}\ee with $p=\frac{a_n}{3},~~ q=\frac{m^2_q}{\sigma}$.

The resulting values of $\omega_0(n), \varepsilon_n (\omega_0)$  are  given
  in   Table 1 for $n=0$, $\sigma=0.18$ GeV$^2$ and  different $m_q$.


\begin{table}
\caption{Energy eigenvalues $\varepsilon_0 (m_q)$ and radial wave function at
origin $R_{IS} (0)$ as functions of quark mass}
\begin{center}
\begin{tabular}{|l|l|l|l|l|l|l|l|l|l|}
\hline\hline$m_q$ (MeV)&0&40&100&200&330&400&500&700&1400\\\hline

$\omega_0 $ ( MeV)& 352&355&373&434&507&578&663&828&1510\\\hline

$\tau(\omega_0) , \alpha_s=0.33$&0.3885&
0.391&0.409&0.447&0.495&0.541&0.5925&0.687&1.026\\\hline $\varepsilon_0$
(MeV)&903&900&879&819&760&711&661&583&380\\\hline $R_{1S} (0)$ Gev$^{3/2}$&
0.312&0.314&0.324&0.358&0.397&0.435&0.457&0.563&0.911\\\hline
\end{tabular}
\end{center}
\end{table}


\begin{table}
\caption{Masses of vector state $M_V$ and pseudoscalar state  $M_{PS}$ together
with spin-averaged mass  $\bar M$ as functions of quark  mass}
\begin{center}
\begin{tabular}{|l|l|l|l|l|l|l|l|l|l|}
\hline\hline$m_q$ (MeV)&0&40&100&200&330&400&500&700&1400\\\hline

$\bar M $ ( MeV)& 766&770&874&1001&1224&1357&1535&1890&3154\\\hline

$M_V$( MeV)&827& 830&932&1053&1271&1401&1572&1926&3182\\\hline $M_{PS}$
(MeV)&458&463&578&733&984&1135&1349&1709&3011\\\hline
\end{tabular}
\end{center}
\end{table}


For large $m_q$, $\frac{m_q^2}{\sigma} \ll 1,$ Eq. (\ref{39}) yields
$\omega^2_0\approx m^2_q +\frac{\sigma^{2/3} a_n m^{2/3}}{3}$.

One can see in Table 1, that $\omega_0$ change only by  3\%, when $m_q$ grows
from 0 to 100 MeV. This fact is  basic  for small $m_q$ corrections to  $f_\pi,
\lan \bar q q\ran$ from the sums oven in (\ref{15}),(\ref{20}).

We now turn to corrections, $\Delta_{SE}$ and $M_{ss}$, which are defined as in
Eq. (\ref{37}), with $R_{1S}(0)$ given
 in the Table 1, and $\eta(m_q\lambda)$ in the Appendix of \cite{i}.

 In this way one obtains for the spin-averaged mass $\bar M$, at  $m_q=0$,

\be \bar{ M }\equiv \bar M_{cog} (1S) =766 {\rm ~MeV},\label{43}\ee and for PS
and $V$ masses with $\alpha_{hf}=0.35 $ and $R_{1S}(0)$ from Table 1 one would
have
$$ M_V = 827 ~{\rm MeV},~~ M_{PS}=583  ~{\rm MeV}.$$ Taking  now $V_{ss}$
interaction to the first order into account for the singlet wave function
$\psi_{PS}(0)$, one has $\psi_{PS}^{(1)} (0) \simeq 1.3 \Psi^{(0)}_{PS} (0)$,
which  shifts $M_{PS}$ down, $M_{PS}^{(!)} = 458$ MeV.

We keep this procedure for all $m_q$ and get in this way the  values of
$M_{PS}$ and $M_v$, given in Table 2.

To make comparison  with lattice data we  demonstrate in Table 3  numerical
values of $M_{PS}, M_V$ and $f_{PS}$, obtained in \cite{f} on large lattices
$16^332$ and $24^348$ at $\beta=8.00$ and $8.45$ respectively. The
corresponding values of $m_q$ were found using the standard procedure with
chiral behavior of $M_{PS}(m_q) \sim \sqrt{m_q}$, as it is also clear from the
last column of Table 3. Comparison of the values of $M_V(m_q)$ in Table 3 and
Table 2 shows an agreement within the accuracy of 10\% \footnote{A few percent
agreement occurs for $\alpha_s\approx 0.4$, but for illustrative purposes we
keep in Tables 1 and 2 $\alpha_s=0.33$.}. The same can be told about the
$M_{PS}$ values for $m_q> m_{\rm crit} \approx 200$ MeV, however for smaller
$m_q$ the RSH calculation predicts $M_{PS}$ slowly changing with a finite value
$O(400$ MeV) for $m_q=0$. This implies that chiral regime is outside of RSH and
around $m_q=200$ MeV a change of regimes takes place.

\begin{table}
\caption{Masses    $M_{PS}$, $M_V$
 and $f_{PS}$ (all in MeV), calculated in \cite{f}, as functions of quark mass.
}
\begin{center}
\begin{tabular}{|l|l|l|l|l|}

\hline \multicolumn{5}{|c|}{$16^332,$ ~~ $\beta=8.00$ }\\\hline

 $m_q$ (MeV)&$M_{PS}$ &$M_V$&$f_{PS}$ &$\frac{m_{PS}}{\sqrt{m_q}}$
Gev$^{1/2}$\\\hline 13.35& 239&809 &94.3&2.068\\\hline
35.2 &295& 816&96.1&1.57\\\hline 52.8&353&829&98.1& 1.53\\
\hline 70.4& 403& 841& 100.6 &1.518\\\hline
 105.6& 488& 874& 103.1& 1.50\\\hline
 176&631&944&113.27& 1.50\\\hline
 246.5& 753&1025&122&1.516\\\hline
 \multicolumn{5}{|c|}{24$^348, ~~~\beta=8.45$}\\\hline
300&887&1286&153&1.62\\\hline 211&731&1159&140&1.59\\\hline
120&551&1036&125&1.59\\\hline

\end{tabular}
\end{center}
\end{table}

One can see in Fig. 3 that our vector masses $M_V$ as a function of $m_q$ with
good accuracy lie on the sequence of two straight lines, and the RSH values of
PS masses are on  an almost parallel line for $m_q>m_{\rm crit}$, while the
chiral branch for $m_q< m_{\rm crit}$ follow the law $M_{PS}^2 \sim m_q$, which
is in agreement with lattice data, shown on Fig.3.

In the next section we shall discuss this change of regimes from another side
of the chiral approach and a possibility of incorporating two regimes into one
scheme.

It is interesting, that the entries in the rightest column of Table 3, are
according to (\ref{27a}), the square root of the gluelump mass, $\sqrt{M_G}$,
which is approximately  equal in Table 3 to 1.5 GeV$^{1/2}, $ and $M_G\approx
O(2$ GeV),  as was predicted in (\ref{27a}),

\section{A universal spectrum in the PS channel}

The corrected GMOR relation (\ref{18}) gives the values of the PS masses in the
wide interval for $m_q<m_{\rm crit}$, where the square root behavior
 $M_{PS} \sim \sqrt{m_q}$ goes over into the  quasilinear RSH regime $M_{PS}\sim a +b m_q$. One might wonder how the quasilinear
regime,present in the spectrum of the RSH Green's function $G^{(0)} (k) =
\sum_n \frac{c^2_n}{k^2+m^2_n}$ is coexisting with the chiral regime  in the
``total" Green's function $G(k)$. To this end  one can compare definitions of
$N(k)$ in (\ref{4}), (\ref{5}) and (\ref{17}) and understand, that $\frac{2
N_c}{f^2_\pi} N(k) = m^2_\pi + k^2 + O(k^4)$  is the inverse of the generalized
pion propagator,   \be \Delta_\pi (k) \equiv  \frac{f^2_\pi}{2N_c N(k)} =
\frac{1}{m^2_\pi + k^2 a(k^2)}\label{37a}\ee where \be a(k^2)
=\frac{2N_cM^2(0)}{f_\pi^2} \sum^\infty_{n=0}\frac{c_n^2}{m^2_n (k^2+m^2_n)}
\equiv \frac{f^2_\pi(k^2)}{f^2_\pi(0)}\label{38a}\ee

Note that $a(0)=1$ due to (\ref{20}).

From (\ref{37a}) one can find poles of $\Delta_\pi(k)$, which are connected to
the poles in $a(k^2)$. The latter are standard  flux-tube eigenstates, the
spin-averaged ones, since $G(k)$ does not contain hyperfine interaction. As a
consequence one obtains for $ m_q=0$

$$
a(k^2):~{\rm poles~ at~} ~~-k^2=m^2_n,   ~~n= 0,1,2,...$$

 \be \Delta_\pi (k):~{\rm poles~ at}~~ k^2=0,;-k^2=m^2_n-\delta m^2_n
~{\rm(chiral)}, ~~n=1,2,...\label{47}\ee

Note, that the $m^2_0$ pole is replaced in $\Delta_\pi (k)$ by the pion pole
(in the chiral limit) $k^2=0$, while $m^2_1$ is shifted down by $\delta m_1^2 $
(chiral).

One can check, that for the initial values $m_0 \simeq 0.5$ GeV, $m_1 =1.3$
GeV, the shift of the mass eigenvalue of $m_\pi (2S)$ is around 0.15 GeV. This
means, that the lowest 1S eigenvalue of the RSH at $m_0$ is replaced by
$m_\pi$, and the 2S state is shifted down by $\sim 10\%$, which is  expected
from the physical considerations, since this shift replaces the hyperfine
interaction.

In the general case of $m_q>0$ the poles of $\Delta_\pi (k)$ are at
$k^2=-M^2_{unif}$ where $M^2_{unif} \approx \frac{m^2_\pi (m_q) m^2_0
(m_q)}{m^2_\pi (m_q) + m^2_0 (m_q)} \left( 1+O\left(
\frac{m^2_\pi(m_q)}{m^2_1(m_q)}\right)\right).$

 The situation with poles is illustrated in   Fig. 4, where poles of
 $\Delta_\pi (k)$  denoted as  $M_{unif}$  are
shown together with lattice values of $M_{Lat}$ and the flux-tube values of
$M_{PS}$, taken from Table 2.  One  can see, that $M_{unif}$ approaches
$M_{PS}$ with growing $m_q$ and all three branches $M_{PS}$, $M_{Lat}$ and
$M_{unif}$  are rather close to each other for $m_q> 150$ MeV. Thus indeed,
unification of chiral and  flux-tube dynamics, automatically obtained on the
lattice  can be paralleled with our  explicit mechanism of unification, given
in (\ref{37a}).

\begin{figure}
\includegraphics[width= 9cm,height=10cm,keepaspectratio=true]{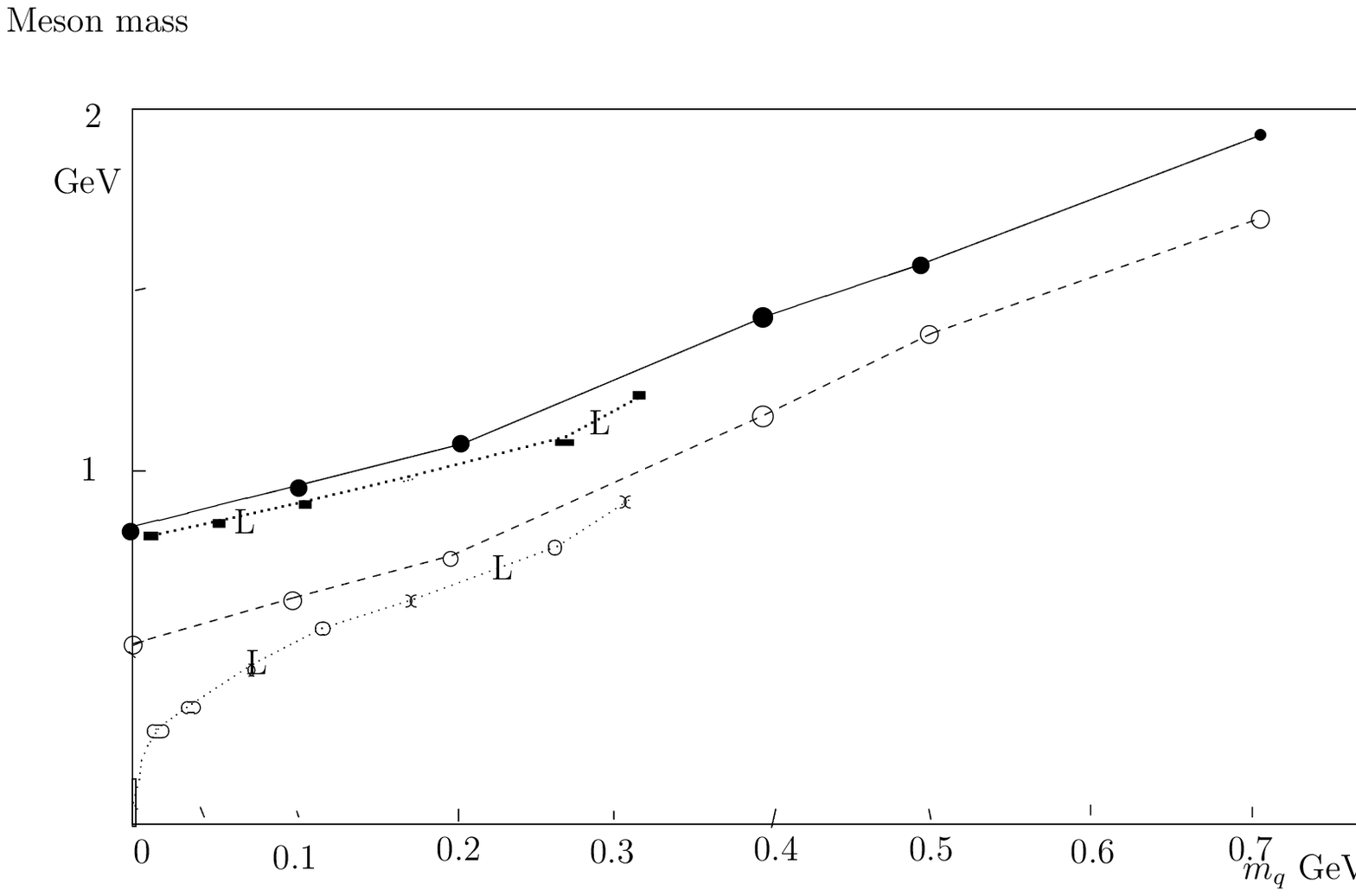}
\caption{{ Spectrum of vector and PS mesons from analytic calculations with the
Hamiltonian (\ref{35}) (solid  and dashed lines respectively) and from lattice
calculations \cite{f} (dotted lines marked by the letter $L$).}} \vspace{3cm}
 \hspace{-6cm}

\end{figure}

\begin{figure}
\includegraphics[width= 9cm,height=10cm,keepaspectratio=true]{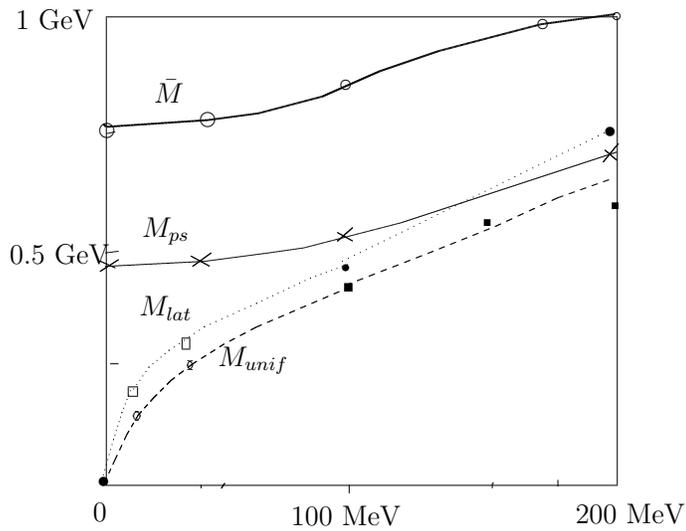}
\caption{Masses of PS mesons as functions of quark mass; $M_{PS}$ calculated in
the flux-tube dynamics (solid  line), $M_{lat}$, calculated on the lattice from
\cite{f}, and $M_{unif}$, as poles of the generalized pion propagator,
Eq.(\ref{37a}). The upper solid line is for the spin-averaged masses $\bar M$,
from  Table 2. } \vspace{1cm}

\end{figure}

\section{Summary and conclusions}

We have studied in the paper above the confining and CSB dynamics using varying
quark masses as a tool to distinguish in the hadron spectra different branches,
corresponding to different mechanisms. Comparison to the existing lattice data
allows to establish a good accuracy of our relativistic formalism, based on
RSH, and an interesting correspondence between the CSB dynamics and the
flux-tube dynamics, and to answer the question, at what quark mass the
Nambu-Goldstone modes transform into the standard flux-tube modes. In section 3
we have found, that it happens approximately at $m_q=m_{crit}\approx 0.2$ GeV.
Another interesting point is that the almost linear dependence of $m_V$ on
$m_q$ slightly changes its slope at  $m'_{crit} = \sqrt{2\sigma} \left(
\frac{a_n}{9}\right)^{3/4} \approx 0.21$ GeV (cf.  Eqs. (\ref{40}) and
(\ref{41})), which again agrees with lattice and analytic calculations, as can
be seen from Fig. 3.

Maybe the most important result of the paper is the improved derivation of the
chiral dynamics directly from the QCD Lagrangian  in section 2,  which enables
one not only  to obtain  GMOR relation, but also  find also  the corresponding
$m_q$ dependencies. This program, started in \cite{a,b,c,d}, allows in
principle  to derive the chiral Lagrangian directly from QCD, including
higher-order corrections, and as was already mentioned above, to connect chiral
and flux-tube dynamics, which are usually treated separately.

Note the important role, which is played in this program by the new entity  --
the vertex (or residual) mass $M(0)$. Its value was calculated in terms of
$\sigma$ and the correlation length $\lambda, M(0) \sim \sigma \lambda$. Note,
that in the flux-tube dynamics, i.e. at distances $\lan r \ran \approx
1/\sqrt{\sigma}$, the  mass $M(0)$ does not  appear, and it is needed only to
calculate $f_\pi$ and $\lan \bar q q\ran$, while in the GMOR relations $M(0)$
is absent.

Looking from the flux-tube dynamics side, it is interesting to try to
understand, which kind of forces causes the breakdown of flux-tube and
appearance of chiral dynamics, and in particular the strong reduction of the
pion mass. To this end consider  in particular, the hyperfine ($h f$)
interaction, deduced in the RSH,  which is singular for small values of
$\omega_i$ (i.e. for small average quark energies), since

$$ M_{ss} (\omega)  =\frac89 \frac{\alpha_{hf}}{\omega}  \sigma (\ves_i\ves_j)$$

Taking this into account and working in the einbein formalism, one can find
minimum of $M(\omega_i =\omega_j)$  in the  PS channel, \be M(\omega) =
\frac{m^2_q}{\omega} - \frac23 \frac{\alpha_{h f}\sigma}{\omega} +
O(\omega^{-1/3}).\label{50}\ee  This minimum exists for $m_q>
m^{\prime\prime}_{crit} = \sqrt{\frac23 \alpha_{h f}\sigma} \approx 0.2~{\rm
GeV}, $ but  disappears for smaller $m_q$,  and $M(\omega)$ is not bounded from
below.

Usually and in the present paper the $h f$ interaction is treated
perturbatively, which in particular means, that the  corresponding $\omega$ in
$M_{h f}$ is taken from the spin-averaged masses $M^{(0)}$.

It is not known   at present how to treat  $M_{hf}$ nonperturbatively for $m_q
< m^{\prime\prime}_{crit}$. It seems very likely, that the divergence  of $M$
for small $\omega$ implies necessity
 of consideration
of multiple pair creation  or, in other words, the reconstruction of the vacuum
with the appearance of quark pair condensate. This  is still another
manifestation of the appearance of the new regime -- the chiral Nambu-Goldstone
regime.

From this point  of view it is probably not surprising, that all three critical
masses $m_{crit}, m'_{crit}, m^{\prime\prime}_{crit}$ approximately coincide.
This fact calls for  further studies  with the aim  of understanding and
unifying chiral and  flux-tube dynamics.

 The author is grateful to  A.M.Badalian  for constant help and useful
 advices.

\vspace{2cm}

{\bf Appendix 1}\\

{\bf Calculation of $f_\pi $ and $\lan \bar q q\ran$}
\\
\setcounter{equation}{0} \def\theequation{A1.\arabic{equation}}

We illustrate in the appendix the methods of section 2
  with an improved calculation of the chiral condensate
and $f_\pi,\lan \bar q  q \ran $ using Eq.(\ref{15}) and (\ref{20})
respectively and the methods, given in \cite{b,c,d}.

Defining these values at $\lambda^{-1}= 2 $ GeV, one should use the
corresponding cut-off factors, and one has\footnote{Extra factors in
(\ref{28}), (\ref{29}), as compared to (\ref{15}), (\ref{20}), are due to
exclusion of a piece of the Euclidean time integration from $0$ to $\lambda$,
see eq. (\ref{11}) of \cite{c}.}

\be - \frac{\lan \bar q q\ran}{n_f} = N_c (M(0) + \bar m_q) \sum ^N_{n=0}
\frac{\psi^2_n (0)}{m_n} e^{-m_n\lambda}\label{28}\ee

\be f^2_\pi = N_c M^2(0) \sum^N_{n=0} \frac{\psi^2_n (0)}{m^3_n}
e^{-m_n\lambda} (1+ m_n \lambda).\label{29}\ee Here $N$ is taken to be 2, since
higher terms are small.

Insertion of $m_n, \psi^2_n(0)$ for $n=0,1,2$ from the  RSH in section 3,  one
obtains

\be - \frac{\lan \bar q q\ran^{(2~{\rm GeV})}}{n_f} = (217~{\rm MeV})^3
\frac{(M(0) +m_q)}{150 ~ {\rm  MeV}}\label{30}\ee

\be  f_\pi^{(2~{\rm GeV})}= 96~{\rm MeV} \frac{(M(0)+m_q)}{150~{\rm
MeV}}.\label{31}\ee

This should be compared with   the  values, obtained on the lattice
\cite{f,g,h,i}, which can be roughly characterized by  an average value in the
quenched case $  - {\lan \bar q q\ran} = [(270\pm 20) ~{\rm MeV}]^3$ while from
\cite{i} for $n_f =2$\be - \frac{\lan \bar q q \ran}{n_f} = (209\pm 8 ~{\rm
MeV})^3.\label{32}\ee

The results (\ref{28}), (\ref{29}) are sensitive to the cut-off factor (vacuum
correlation length) $\lambda$, e.g. for $\lambda= 1.5$ GeV$^{-1}$ one obtains

\be - \frac{\lan \bar q q\ran^{1.5~{\rm GeV}}}{n_f} = (195~{\rm MeV})^3
\frac{(M(0) +m_q)}{120 ~ {\rm  MeV}}\label{33}\ee

\be  f_\pi^{1.5~{\rm GeV}} =64 ~{\rm MeV} \left(\frac{(M(0)+m_q)}{120~{\rm
MeV}}\right).\label{34}\ee

In the paper  we  adopt the results (\ref{30}), (\ref{31}) and  the
corresponding estimate $M(0) =0.15$ GeV as our current values.

\end{document}